\documentclass{jfm}

\usepackage{graphicx}
\usepackage{newtxtext}
\usepackage{newtxmath}
\usepackage{natbib}
\usepackage{hyperref}
\usepackage{multirow}
\hypersetup{
    colorlinks = true,
    urlcolor   = blue,
    citecolor  = blue, 
}

\newcommand{\RomanNumeralCaps}[1]
\linenumbers

\title{Multiscale circulation in wall-parallel planes of turbulent channel flows}

\author{Peng-Yu Duan\aff{1},
  Xi Chen\aff{1} \corresp{\email{chenxi97@outlook.com}}
 \and Katepalli R. Sreenivasan \aff{2} }

\affiliation{
\aff{1}Key Laboratory of Fluid Mechanics of Ministry of Education, Beihang University (Beijing University of Aeronautics and Astronautics), Beijing, 100191, PR China
\aff{2}Tandon School of Engineering, Courant Institute of Mathematical Sciences, and Department of Physics, New York University, New York 10012, USA
}

\begin{document}
\maketitle

\begin{abstract}
Wall turbulence consists of various sizes of vortical structures that induce flow circulation around a wide range of closed Eulerian loops. Here we investigate the multiscale properties of circulation around such loops in statistically homogeneous planes parallel to the wall.
Using a high-resolution direct numerical simulation database of turbulent channels at Reynolds numbers of $Re_\tau=180$, 550, 1000 and 5200, circulation statistics are obtained in planes at different wall-normal heights. Intermittency of circulation in the planes of the outer flow ($y^+ \gtrsim 0.1Re_\tau$) takes the form of universal bifractality as in homogeneous and isotropic turbulence. The bifractal character simplifies to space-filling character close to the wall, with scaling exponents that are linear in the moment order, and lower than those given by the Kolmogorov paradigm. The probability density functions of circulation are long-tailed in the outer bifractal region, with evidence showing their invariance with respect to the loop aspect ratio, while those in the inner region are closely Gaussian. The unifractality near the wall implies that the circulation there is not intermittent in character.



\end{abstract}


\section{Introduction}
\label{sec:intro}


The most important diagnostic quantity for characterizing the inertial-range intermittency in turbulence is the velocity increment over a separation distance contained in the inertial range (IR). This practice has been used for more than 80 years since Kolmogorov's pioneering work of 1941 (K41 henceforth), and has produced a quantitative understanding of the scaling exponents and their multifractal modeling (see, e.g., \cite{ Frisch_turbulence, Sreeni1997_ARFM_turb}). Years later, \cite{Migdal94_area_rule,Migdal2023} proposed that circulation around Eulerian loops of various sizes makes a more natural connection to fluid mechanics. Specifically, he considered
the circulation around a loop of area $A$ defined as
\begin{equation}
\label{eq:def_circ}
    \Gamma_A=\oint_{C}{\boldsymbol{u}^\prime\cdot d\boldsymbol{l}}=\iint_{A}{\boldsymbol{\omega}\cdot\boldsymbol{n}dA},
\end{equation}
where $C$ is the boundary of a loop of area $A$, $\boldsymbol{u^\prime}$ is the fluctuating velocity, $d\boldsymbol{l}$ is an elemental length along $C$, $\boldsymbol{\omega}=\boldsymbol{\nabla}\times\boldsymbol{u}^\prime$ is the fluctuating vorticity, and $\boldsymbol{n}dA$ is an elemental area of $A$ in the direction of the unit normal $\boldsymbol{n}$. It is trivial to show that the circulation moments scale as $\langle\left|\Gamma_A\right|^p\rangle \sim A^{2p/3}$ in Kolmogorov's 1941 paradigm. However, intermittency introduces deviations from this scaling, requiring a more general power-law form
\begin{equation}
\langle\left|\Gamma_A\right|^p\rangle \sim A^{\zeta_p/2} \sim r^{\zeta_p},
\end{equation}
where $r \sim \sqrt A$ is the linear dimension of the loop.
Absolute values are used in (1.2) because odd moments of $\Gamma_A$ cancel out due to the symmetry of the PDF. \cite{Migdal94_area_rule} proposed the area rule that the tails of the probability density functions (PDF) of $\Gamma_A$ depend only on the minimal area of the loop $C$ in IR, not on the shape of the loop. Early work \citep{Sreeni1995_circ,Cao96_PRL,Benzi97_PRE_circ_in_shear} attempted to make connections with the theory, but it was hampered by the low Reynolds numbers of the flows.
The stimulating work of \cite{ISY19} in homogeneous and isotropic turbulence (HIT) at high Reynolds numbers has shown that a concise bifractal relation holds for $\zeta_p$, thus revealing considerable simplicity in the intermittent structure of circulation. This result is plausible because circulation, being the area integral of vorticity, would smooth out extreme local fluctuations of velocity gradients. \cite{Iyer21_PNAS} also confirmed the area rule for the PDFs. Since then,
\cite{Muller21_PRX_quantum} and \cite{Muller21_NC_quantum} demonstrated that the circulation in quantum turbulence is also a bifractal, linking the intermittency of quantum and classical turbulence.
Similarly, \cite{Zhu23_PRL_2D_circ} confirmed the bifractality of circulation in the inertial range of the inverse energy cascade of quasi-two-dimensional turbulence experiments. \cite{Muller24_PRL_quantum_to_2D} {compared this result with those in incompressible quantum turbulence and found the equivalence of circulation intermittency in the two instances.}

These studies have focused primarily on HIT. The early investigations of circulation by \cite{Sreeni1995_circ} were made in wakes, and those of \cite{Benzi97_PRE_circ_in_shear} in periodically varying shear flows, but, again, the Reynolds number of these studies were small, as was the shear. Recently, \cite{KAUST_circ_JoT} and \cite{KAUST_PoF_2024} examined the evolution of circulation in turbulent flow that passes through an experimental contraction subjected to mean strain, and found an approximate bifractality of circulation, corresponding to the results of \cite{ISY19}.
However, in none of these flows was the variation of the mean shear strong, and the viscosity effects as central, as in turbulent channel flows. Filling this gap is the central purpose of this paper.

Here, we study the statistics of circulation in wall-parallel planes of turbulent channel flows. We have three particular reasons for these studies. First, wall turbulence is characterized by a rich set of coherent structures \citep{Kline1967_coherent_struc,Adrian2000,Jimenez_2018} spanning from small to large to very large scales. These structures, particularly vortical structures in different orientations, can produce an effect on circulation, deserving a quantitative investigation of their multiscale properties. Second, with increasing height from the wall, the shear and anisotropic effects tend to vanish, and hence the circulation in the homogeneous center plane may be expected to be similar to that in HIT. In other words, it would be interesting to examine how the fractality of circulation statistics varies with the height. Finally, the Reynolds number variation of circulation properties enriches the understanding of Reynolds number similarity and offers insights for developing eddy-based turbulent models.




\section{Data for analysis}
\label{sec:data}

\begin{figure}
    \centering
    \includegraphics[width=1.0\linewidth]{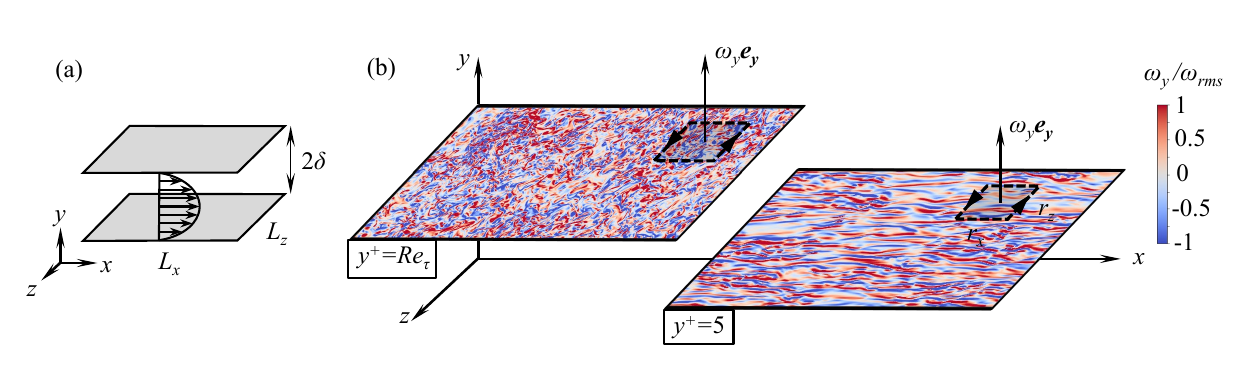}
    \caption{(a) Sketch of the computational domain for channel where $L_x$ and $L_z$ are the streamwise and spanwise sizes, respectively. (b) An illustration of the circulation around a rectangular loop with side lengths $r_x$ and $r_z$ in a wall-parallel plane, top-left for $y^+=Re_\tau$ and bottom-right for $y^+=5$ (superscript '+' denotes normalization in wall units). The panels show contours of the normalized wall-normal vorticity $\omega_y\boldsymbol{e_y}$ in the domain of 2000$\eta$ in length and 1000$\eta$ in width using the DNS data at $Re_\tau=5200$. Spot-like structures appear in the center plane while streak-like structures are visible near the wall.}
    \label{fig:sim_config}
\end{figure}

\begin{table}
    \centering
    \begin{tabular}{cccccccc}
$Re_\tau$ & $L_x/\delta$, $L_z/\delta$ & $N_x,N_y,N_z$ & $\Delta x/\eta$ & $\Delta z/\eta$ & $\Delta y/\eta$ & $L_x/\eta$ & $L_z/\eta$\\
180	& 4$\pi$, 2$\pi$ & 256, 128, 128 & 2.3-6.2 & 2.3-6.2 & 0.3-1.6 & 585-1589 & 293-794 \\
550	& 4$\pi$, 2$\pi$ & 512, 256, 256 & 2.6-10.6 & 2.6-10.6 & 0.5-1.9 & 1338-5437 & 669-2718 \\
1000 & 8$\pi$, 3$\pi$ & 2048, 512, 1536 & 2.6-10.3 & 1.3-5.2 & 0.002-1.5 & 5363-21181 & 2681-10590 \\
5200 & 8$\pi$, 3$\pi$ & 10240, 1536, 7680 & 1.8-11.1 & 0.9-5.5 & 0.06-1.7 & 18635-113635 & 6988-42613
    \end{tabular}
    \caption{Discretization of direct numerical simulations of Navier-Stokes equation. Cases of $Re_\tau=180$ and 550 are our current simulations, and $Re_\tau=1000$ and 5200 are simulations by \cite{LM2015} from Johns Hopkins turbulence database. Note that $\Delta x$ and $\Delta z$ are uniform grids, while $\Delta y$ stretches in the wall-normal direction in terms of $\eta$, which is the local Kolmogorov length scale.}
    \label{tab:config}
\end{table}

DNS data for turbulent channels used in this paper are for $Re_\tau=180$, 550, 1000, and 5200. Here, $Re_\tau=u_\tau\delta/\nu$, with $u_\tau=\sqrt{\tau_w/\rho}$ as friction velocity, $\tau_w$ the mean wall shear stress, $\rho$ the density, $\nu$ the viscosity and $\delta$ the half height of the channel. Our data for the first two Reynolds numbers have been validated in \cite{Jiabin21}. For $Re_\tau=1000$ and 5200, the data have been simulated by \cite{LM2015}, available from the Johns Hopkins Turbulence Database \citep{JHU_database}. 

Table \ref{tab:config} shows the grid spacings in terms of the Kolmogorov length scale $\eta=(\nu^3 / \langle\epsilon\rangle)^{1/4}$, where $\langle\epsilon\rangle = \langle\nu\partial_j u_i^\prime\partial_j u_i^\prime \rangle$ is the local turbulent dissipation rate. Here and elsewhere, $\langle \cdot \rangle$ denotes the ensemble average. By nominal standards, this resolution ensures that small scales are resolved. As the dissipation decreases with the wall-normal distance $y$, $\eta$ increases from its minimum at the wall to its maximum at the center, which is also why the grid spacings normalized by $\eta$ vary from the wall to the center.

Fully developed turbulent channels are homogeneous in wall-parallel $x$-$z$ planes (figure \ref{fig:sim_config}a), so that circulation in those planes is the focus here. This also enables a potential comparison with the results obtained from HIT. A sketch of the circulation around a rectangular loop in the $x$-$z$ plane is shown in figure \ref{fig:sim_config}(b) for $y^+=Re_\tau$, i.e. the center plane, and for $y^+=5$, a plane in the viscous sublayer. The former plane features vanishing mean shear with blob-like vorticity, whereas the latter experiences strong mean shear with dominant coherent structures, such as velocity streaks \citep{Kline1967_coherent_struc}. We thus expect a notable difference in circulation characteristics for planes with increasing wall-normal distance.

Experiments typically use loop integration as they provide more accurate velocity measurements than velocity gradients \citep{Sreeni1995_circ,KAUST_circ_JoT}, but we calculate the circulation from DNS data via vorticity integration in (\ref{eq:def_circ}). The integrated area is denoted as $A=r_x\times r_z$ and $\omega_y \boldsymbol{e}_y$ indicates the wall-normal vorticity. Both methods were attempted in \cite{Iyer21_PNAS}, obtaining complete equivalence except for loops with side length of $\eta$. Square loops are first considered, i.e. $r_x=r_z$ as in figure \ref{fig:sim_config}(b), while the influence of loop shape will be discussed in section 3. As shown in table \ref{tab:config}, the sampling area for averaging at $Re_\tau=5200$ spans from $60\eta^2$ to $10^8\eta^2$ in the center and from $2\eta^2$ to $10^9\eta^2$ near the wall, thus offering a wide scale range for examining scaling properties. For $Re_\tau=180$, the sampling area extends from $5\eta^2$ to $10^5\eta^2$ in the center; although smaller compared to $Re_\tau=5200$, it is sufficient to verify the scaling by extended self-similarity \citep{Benzi97_PRE_circ_in_shear}, as described in section 4.

\section{Scaling in the inertial range for different wall-normal heights}
\label{sec:results}

\begin{figure}
    \centering
    \includegraphics[width=1.0\textwidth]{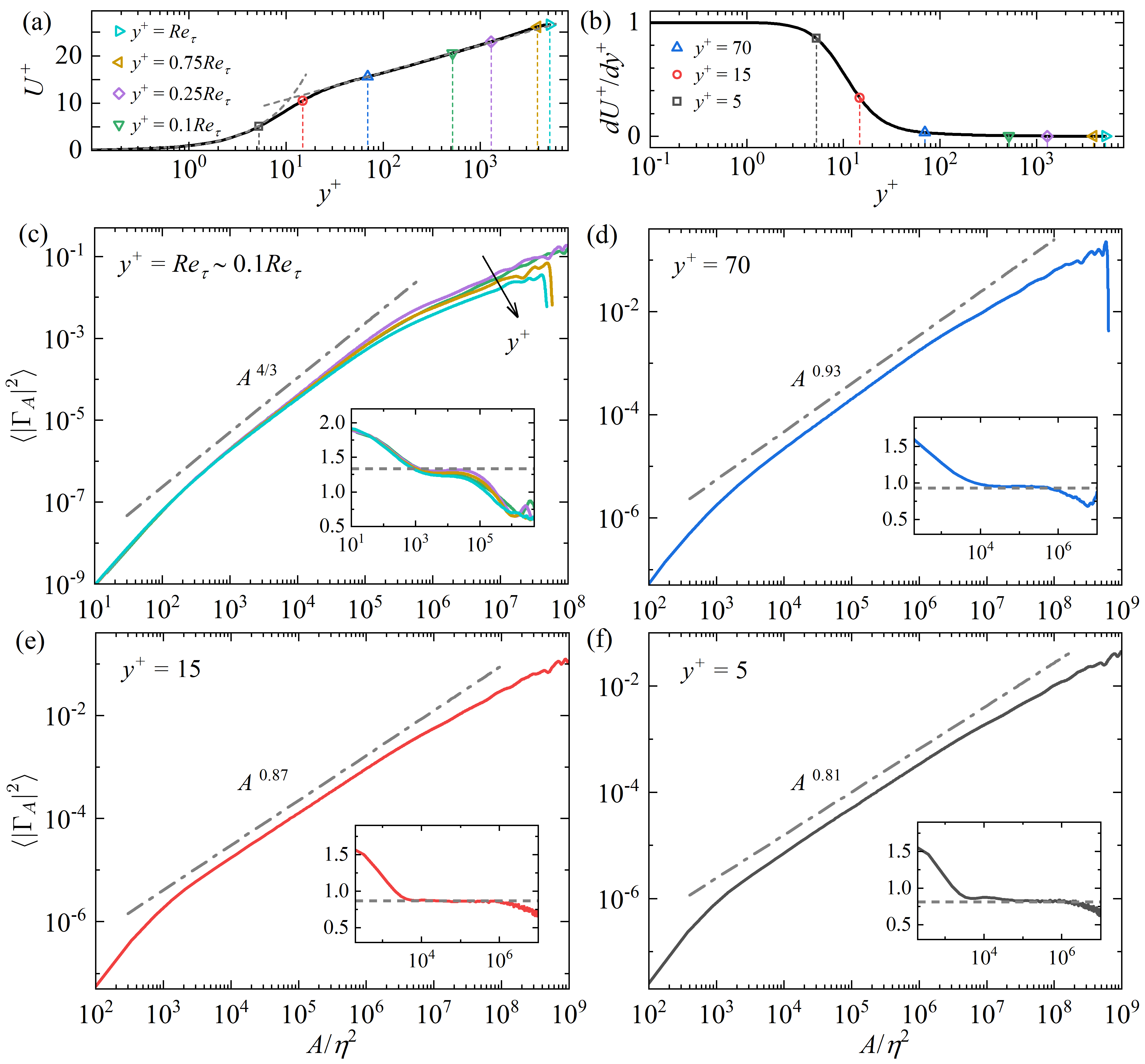}
    \caption{(a) Mean velocity profile and (b) mean shear distribution for $Re_\tau=5200$, with dots highlighting the selected heights. The discrete points marked on the profiles correspond to planes on which circulation statistics are presented. Their colors carry over to figure 3.
    (c-f) Second-order circulation moments (or the variance of circulation) as a function of the loop area $A/\eta^2$, with line colors corresponding to the selected heights in panels (a) and (b). Inset shows the corresponding local exponents, with dash-dotted line indicating the scaling in the inertial range for each case.}
    \label{fig:mean_u_and_2nd_5200_heights}
\end{figure}

\begin{figure}
    \centering
    \includegraphics[width=1.0\textwidth]{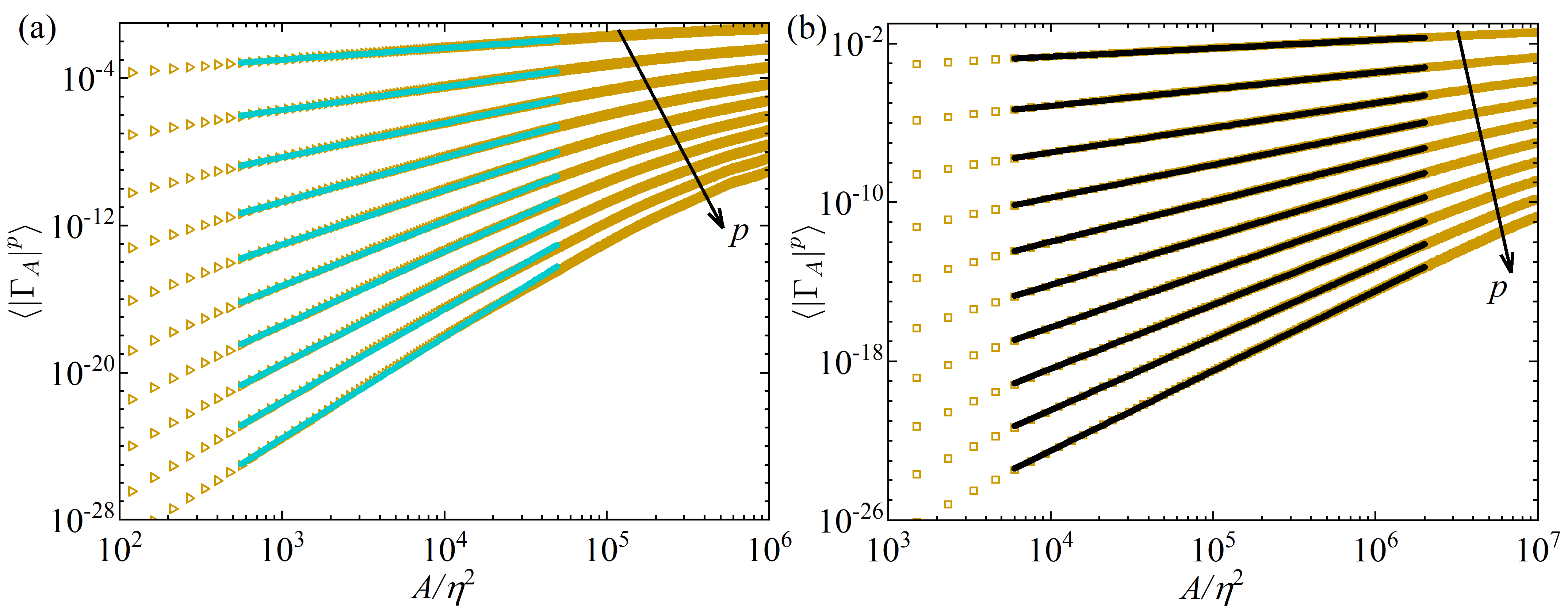}
    \caption{Determination of power-law scaling exponents by fitting circulation moments within the inertial range at (a) $y^+=Re_\tau$ and (b) $y^+=5$. Symbols are simulation data at $Re_\tau=5200$, and solid lines (colors corresponding to the positions depicted in figures 2(a,b)) are the fitting results.}
    \label{fig:5200_IR_fit}
\end{figure}

We first focus on the case of $Re_\tau=5200$ for which an inertial range is clearly present. To capture circulation properties at different shear levels, seven heights are selected, i.e.
$y^+=5200~(Re_\tau)$, $3900~(0.75Re_\tau)$, $1300~(0.25Re_\tau)$, $520~(0.1Re_\tau$), $70$, 15 and 5, marked by different symbols in figures \ref{fig:mean_u_and_2nd_5200_heights}(a,b).
In the outer region (including the so-called logarithmic layer and the core), where $y^+\gtrsim 0.1Re_\tau$, the mean shear is negligible; while in the inner region, the flow is shear dominated (figure \ref{fig:mean_u_and_2nd_5200_heights}b) and highly anisotropic. The current dimensionless parameter $S^+=S\nu/u_\tau^2=dU^+/dy^+$ is equivalent to the definition $S^*=Sk/\epsilon$, which has been used, among others, most recently by \cite{KAUST_PoF_2024}, if one adopts the classical wall scaling for the kinetic energy to be $k \sim u_\tau^2$, and for the dissipation rate to be $\epsilon \sim u_\tau^4/\nu$. According to the DNS data used here, we find $S^*>20$ for $y^+<20$, while $S^*<10$ above the buffer layer.

The circulation variances at these selected heights are displayed in figures \ref{fig:mean_u_and_2nd_5200_heights}(c-f), all exhibiting unambiguous scaling. The insets further show the local slopes, whose plateau region extends for approximately two decades of IR in the loop area. In particular for the four heights in the outer region (figure \ref{fig:mean_u_and_2nd_5200_heights}c), all curves collapse on each other and closely correspond to the Kolmogorov scaling of $A^{4/3}$, quite similar to that in HIT. This concurrence indicates that the mean shear effect is, in fact, negligible for the outer circulation. In contrast, closer to the wall, the scaling exponents are not universal, gradually decreasing with smaller $y^+$: $A^{0.93}$ at $y^+=70$, $A^{0.87}$ at $y^+=15$ and $A^{0.81}$ at $y^+=5$.

\begin{figure}
    \centering
    \includegraphics[scale=1.0]{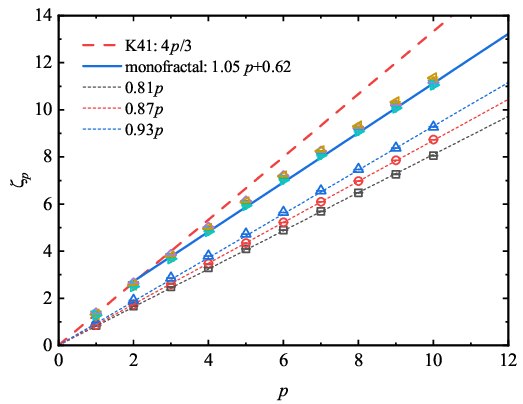}
    \caption{Inertial-range scaling exponents $\zeta_p$ as a function of moment order $p$ at different heights for $Re_\tau=5200$. The long-dashed line is the K41 prediction, $\zeta_p=4p/3$. The solid line indicates the unifractal model for results above the log layer, while the short-dashed lines passing through the origin indicates unifractality below the log layer.
    Symbols are consistent with figures \ref{fig:mean_u_and_2nd_5200_heights}(a-b). }
    \label{fig:scaling_exp_heights_5200}
\end{figure}

By extending the findings of figure \ref{fig:mean_u_and_2nd_5200_heights} to higher orders, we can assess intermittency effects under varying shear levels. High-order moments are presented in figures \ref{fig:5200_IR_fit}(a) and \ref{fig:5200_IR_fit}(b) for the two planes of $y^+=Re_\tau$ and $y^+=5$, respectively. The best power-laws in the inertial range, marked by solid lines, are in close agreement with the data. We repeat this procedure for all the selected planes and collect in figure \ref{fig:scaling_exp_heights_5200} the scaling exponents of circulation moments for orders $p=1-10$, with statistical uncertainties (obtained using the student's \textit{t}-distribution with 95\% confidence intervals) subsumed by symbol thicknesses.

Two notable observations should be made about figure \ref{fig:scaling_exp_heights_5200}. First, in the inner layer ($y^+\lesssim 70$) the data for all moment orders are best fitted by straight lines $\zeta_p = k_{y}p$ without intercepts, where $k_y$ increases with wall distance. This property implies that  circulation in the inner planes resides on space-filling unifractal sets. Not surprisingly, the slopes of unifractality $k_y$ in figure \ref{fig:scaling_exp_heights_5200} are identical to the scaling exponents of their variance $\langle\left|\Gamma_A\right|^2\rangle\sim A^{k_y}$ in figures \ref{fig:mean_u_and_2nd_5200_heights}.

The second observation is that circulation in outer planes ($y^+\gtrsim0.1Re_\tau$) exhibits a bifractal behavior. A single linear relation with zero intercept cannot be fitted to the data but can be approximated well by two straight lines, one for $p<2$ the other for $p \ge 2$. The former is consistent with K41, while the latter can be fitted by the relation
\begin{equation}
\label{eq:zeta_p}
    \zeta_p = hp+(2-D),
\end{equation}
 where $h$ is the H\"older exponent representing the degree of singularity and $D$ represents the corresponding fractal dimension. Note that the Hölder exponent quantifies degree of local singularity of a relevant physical quantity, with the smaller $h$ signifying a stronger singularity, see, e.g., \cite{Frisch_turbulence}. Impressively, both $h=1.05$ and the dimension $D=1.38$ are invariant with respect to the wall-normal height in the outer region, suggesting the universality of circulation in this region.

In summary, circulation in the outer region resides on a single bifractal set with known H\"older exponent as well as dimension, similar to HIT. On the other hand, circulation in the inner layer resides in sub-K41 unifractal sets whose dimension varies with the height of the wall. The transition to the outer bifractal behavior appears quite gradual. The uniform bifractal behavior in the outer region compared to the unifractality in the inner region highlights the influence on flow structures from different levels of mean shear.


\begin{figure}
    \centering
    \includegraphics[width=1.0\linewidth]{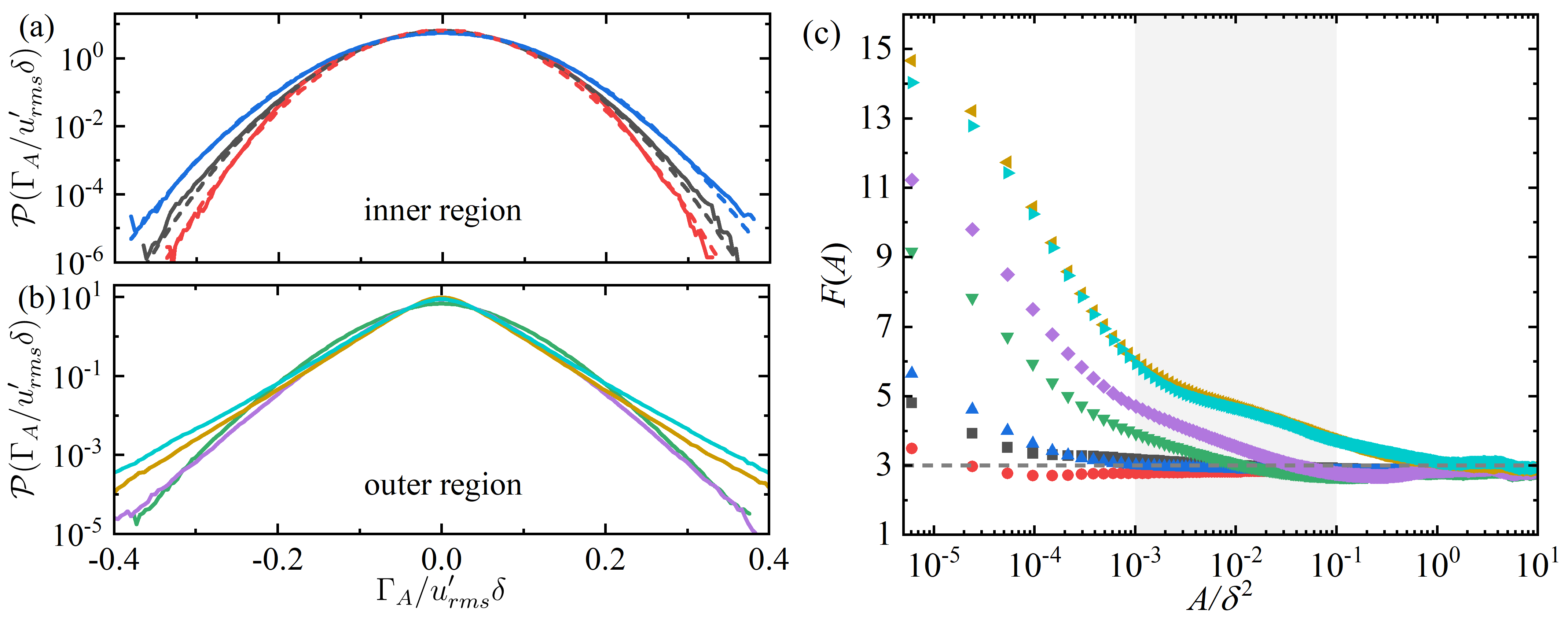}
    \caption{Normalized probability density function $\mathcal{P}$ of $\Gamma_A/(u^\prime_{rms}\delta)$ in (a) inner region (with dashed lines representing the Gaussian distribution) and (b) outer region, both with a loop size of $A/\delta^2=0.0036$ in the IRs.
    (c) Circulation flatness $F(A)$ at different heights as a function of loop size for $Re_\tau=5200$; dashed line indicates the Gaussian flatness of 3. The gray shading highlights the inertial range $A/\delta^2\sim 10^{-3}$ to $10^{-1}$.
    Symbols and line colors are consistent with figures \ref{fig:mean_u_and_2nd_5200_heights}(a-b).}
    \label{fig:pdf_and_flatness}
\end{figure}

We now turn to the PDFs of circulation. Clearly, as shown in figure \ref{fig:pdf_and_flatness}(a), the PDFs are essentially Gaussian in the inner region. This is also reflected in the circulation flatness $F(A)=\langle\Gamma_A^4\rangle / \langle\Gamma_A^2\rangle^2$, which is about $3$ in the scaling region represented by the shaded region in figure \ref{fig:pdf_and_flatness}(c).
Based on the Gaussian PDF, one readily has the relation between $p$-th order moments and the variance: $\langle \left|\Gamma_A\right| ^p\rangle \propto \langle \left|\Gamma_A\right| ^2\rangle^{p/2}$. Because $\langle \left|\Gamma_A\right|^2\rangle\sim A^{k_y}$ has been validated in figure \ref{fig:mean_u_and_2nd_5200_heights}, we have $\langle \left|\Gamma_A\right|^p\rangle \sim A^{ k_y p /2}$, and thus $\zeta_p=k_y p$.
In contrast, the PDFs in the outer region depart strongly from the Gaussian distribution, with approximately stretched exponential functions (figure \ref{fig:pdf_and_flatness}b), so a bifractal intermittency appears. The difference between inner and outer regions could be attributed to viscous effect, as it can damp out extreme events and suppress intermittency, leading to Gaussian PDFs near the wall. Also note in figure \ref{fig:pdf_and_flatness}(a) the non-monotonic PDFs at $y^+=5$, $15$ and $70$. These might be due to the non-monotonic variation of $u'_{rms}$, which has its maximum at $y^+ = 15$. Such a maximum corresponds to prominent velocity streaks, which are induced by wall-normal motions (e.g. sweeps and ejections) of streamwise vorticies, so that circulation in wall-parallel planes is less intense and hence has narrower PDF at $y^+ = 15$.

\begin{figure}
    \centering
    \includegraphics[width=1.0\linewidth]{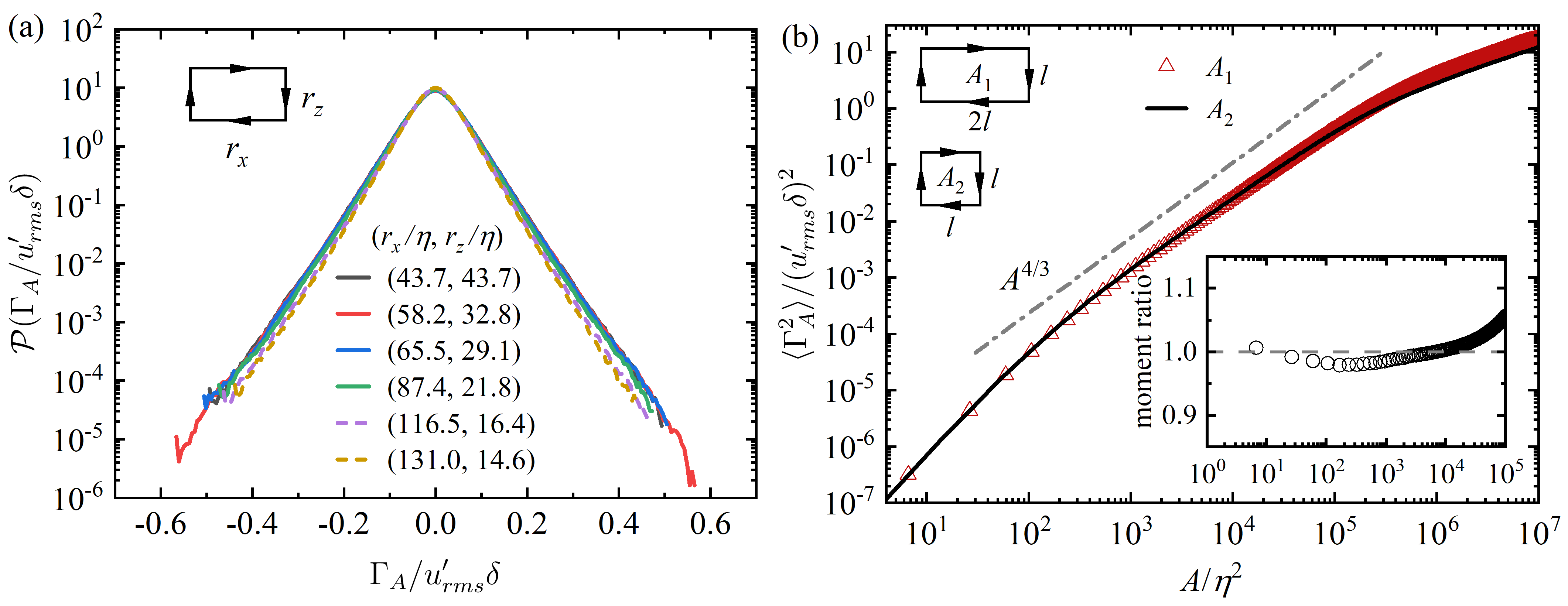}
    \caption{(a) Normalized probability density function of normalized circulation around closed loops with a fixed area but varying aspect ratios. Both (a) and (b) are sampling at the channel center-plane for $Re_\tau=5200$. Solid lines correspond to loops with both sides contained within the inertial region; they collapse on each other. Dashed lines indicate loops with one side outside the inertial range, and thus depart from the collapsed curves. (b) Normalized second-order moments of circulation as a function of area. Data for loop ratio $2:1$ are represented by symbols, and for $1:1$ by solid line. The dash-dotted line indicates the K41’s scaling $\langle \Gamma_A^2 \rangle\sim A^{4/3}$. Inset shows the relative difference for data for the two loop ratios; it is close to unity shown by the dashed line.  }
    \label{fig:area_pdf_and_moments}
\end{figure}

The data considered so far are for square loops. We now explore the impact of the aspect ratio of rectangular loops to assess the area rule, which states that the circulation properties depend only on the loop area \citep{Migdal94_area_rule}. This rule was verified conclusively by \cite{ISY19,Iyer21_PNAS} in HIT at high Reynolds numbers, and somewhat tentatively because of the low Reynolds numbers by \cite{Cao96_PRL} in HIT and by \cite{Benzi97_PRE_circ_in_shear} in shear flows. For the present case, figure \ref{fig:area_pdf_and_moments}(a) shows the PDFs of $\Gamma_A$ for six rectangular loops with the same area $A=1908\eta^2$, but varying aspect ratios $r_x : r_y$ from $1:1$ to $9:1$. The PDFs collapse well when both sides of the rectangle lie within IR but, as expected, deviations occur when one side of the loop extends outside IR (for loops with $r_x/\eta>100$). Furthermore, figure \ref{fig:area_pdf_and_moments}(b) shows the circulation variance as a function of area for two different aspect ratios, that is, $r_x : r_y=1:1$ and $1:2$. In general, the two curves collapse well with each other, particularly within IR ($A/\eta^2 \lesssim 10^4$). Indeed, as shown in the inset of figure \ref{fig:area_pdf_and_moments}(b), the differences between two data sets are less than 2\%. These plots provide clear evidence that circulation properties depend only on the loop area instead of its shape. This is an important conclusion as it suggests the existence of dynamical invariance with respect to variously shaped vortical structures such as hairpins, horseshoes and vortex-packets.


\section{Extended self-similarity at moderate Reynolds numbers}
\label{sec:ESS}

\begin{figure}
    \centering
    \includegraphics[width=1.0\linewidth]{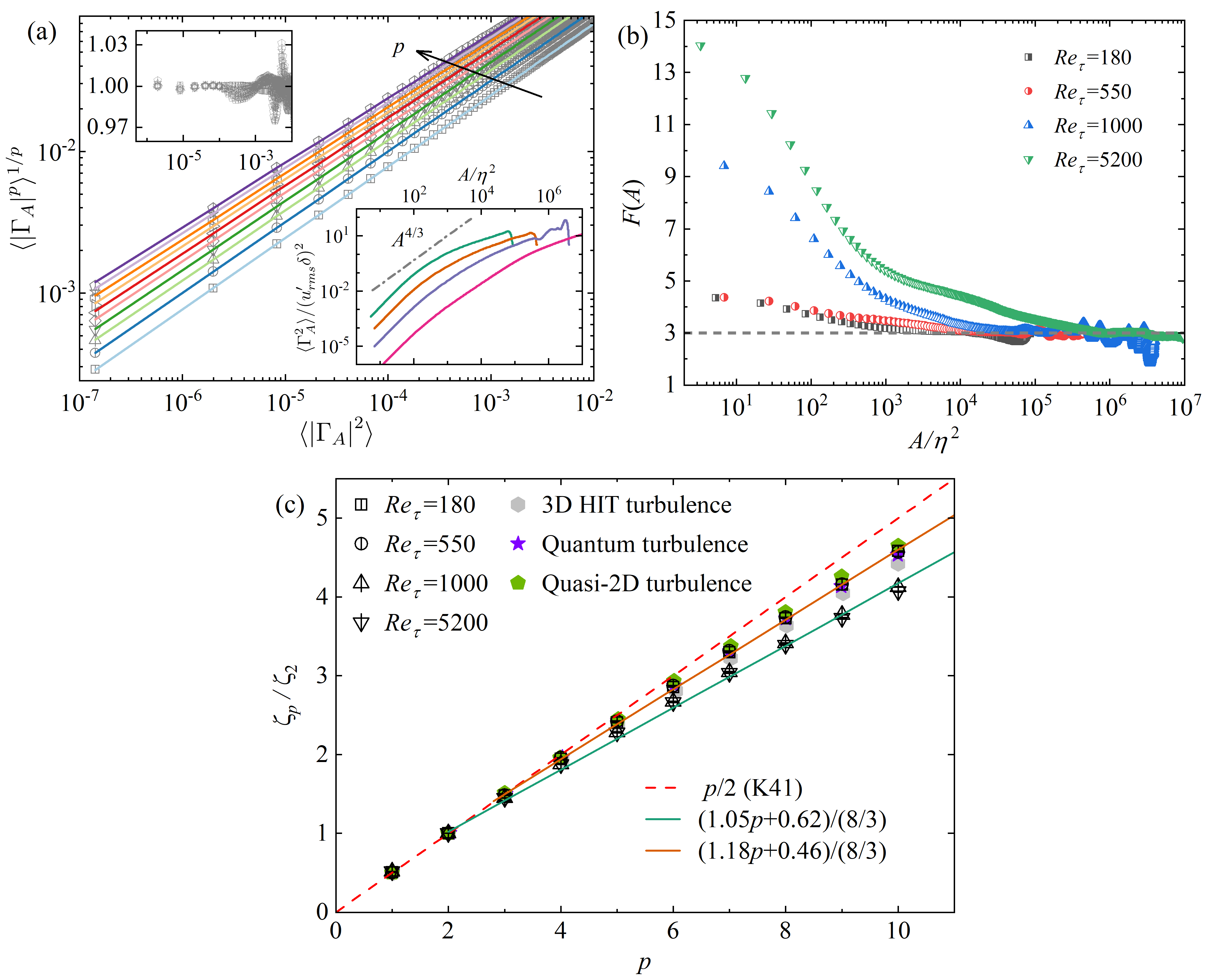}
    \caption{(a) ESS plot of  $\langle\left|\Gamma_A\right|^p\rangle^{1/p}$ versus $\langle\left|\Gamma_A\right|^2\rangle$ at the channel center for $Re_\tau=550$. Symbols are data, and lines are the power-law fits. The top inset shows the relative difference between the  fits and data; the bottom inset shows the center-plane circulation variance for $Re_\tau=180$, 550, 1000, and 5200 (from top to bottom), with the dash-dotted lines representing the K41's scaling.
    (b) Circulation flatness at the channel center for all $Re_\tau$ cases (symbols) compared with the Gaussian value $3$ (dashed line).
    (c) ESS scaling exponents $\zeta_p/\zeta_2$ at the channel center for all $Re_\tau$'s. The dashed line is K41, and the solid lines are the monofractal fits. Data of three-dimensional HIT \citep{ISY19}, quantum turbulence \citep{Muller21_PRX_quantum} and quasi-two-dimensional turbulence \citep{Zhu23_PRL_2D_circ} are also included for comparison.}
    \label{fig:ESS}
\end{figure}



We now investigate whether the bifractality of the circulation in the center plane persists for lower Reynolds numbers. The circulation moments are shown in figure \ref{fig:ESS}(a), with the flatness at different Reynolds numbers shown in figure \ref{fig:ESS}(b), and the relative scaling exponents shown in figure \ref{fig:ESS}(c). Several points are made below.

First, from the bottom inset of figure \ref{fig:ESS}(a), there is no clear inertial range for $Re_\tau \leq 1000$. But ESS \citep{Benzi97_PRE_circ_in_shear} extends the scaling regime, i.e. $\langle\left|\Gamma_A\right|^p\rangle^{1/p}$ versus $\langle\left|\Gamma_A\right|^2\rangle$, as shown in the main panel of figure \ref{fig:ESS}(a) for $Re_\tau=550$.
The extended scaling range enables a robust determination of the relative scaling exponents as the power law over several decades of the new abscissa. The least square fit by a power law achieves errors within 3\% over four decades of $\langle\left|\Gamma_A\right|^2\rangle$, shown in the top inset of figure \ref{fig:ESS}(a). We have also examined the ESS scaling for $Re_\tau=5200$, which is closely in agreement with direct measurement in the inertial range, thus validating the reliability of ESS.

Second, we note that the ESS of $Re_\tau=180$ and 550 are closer to the slope of K41 compared to higher $Re_\tau$, whose explanations could be found in figure \ref{fig:ESS}(b). That is, for small $Re_\tau$, the viscous effect is stronger, and the flatness is closer to the Gaussian value $F(A)=3$, and thus the intermittency of circulation is weaker. However, compared to the Gaussian value, there are still deviations in the inertial range, e.g. $F(A)\approx 4.5$ at $A/\eta^2\sim10^2$, so bifractality is maintained even for $Re_\tau=180$ and 550.

Finally, the last point of the previous paragraph is expanded here. The relative scaling exponents shown in figure \ref{fig:ESS}(c) depart from the K41 paradigm $\zeta_p/\zeta_2=p/2$ for high order $p$.
Similar to the linear relation of (\ref{eq:zeta_p}), the relative exponents for $p>2$ can be fitted by the straight line
\begin{equation}
    \zeta_p/\zeta_2 = [hp+(2-D)]/\zeta_2,
\end{equation}
where $\zeta_2 = 8/3$ by taking K41's scaling as a common normalization (it does not necessarily mean that K41 is valid for low $Re_\tau$). For $Re_\tau=1000$ and 5200, the data collapse with the same $h=1.05$ and $D=1.38$ as shown in figure \ref{fig:scaling_exp_heights_5200}, indicating an asymptotical $Re_\tau$-invariance. For $Re_\tau=180$ and 550, they are higher than those of $Re_\tau \gtrsim 1000$ with $h=1.18$ and $D=1.54$, closer to the observation of HIT \citep{ISY19}, quasi-two-dimensional turbulence \citep{Zhu23_PRL_2D_circ} and quantum turbulence \citep{Muller21_NC_quantum, Muller24_PRL_quantum_to_2D}. Therefore, the bifractality of circulation is a general property of turbulence despite differences in geometries and Reynolds numbers of turbulent structures.

Note that for lower Reynolds numbers ($Re_\tau\leq1000$), the verification of area rule is not performed here, as the inertial range has not been well developed. Also, in planes that are not parallel to the wall, due to the inhomogeneous and anisotropic effect, the shape and orientation of Eulerian loops may affect the statistics of circulation and deserve future studies.

\section{Conclusion}
We have demonstrated that circulation in wall-parallel planes of wall turbulence is a bifractal as long as the height is above the log layer, irrespective of the Reynolds number; this is in agreement with the findings of \cite{ISY19} in HIT. Below the log layer, the bifractality diminishes, transitioning to a space-filling behavior with a scaling exponent lower than the K41 prediction. In particular, the bifractal parameters in the outer region are essentially independent of the height, suggesting a uniform geometric feature of wall-normal vorticity, and the area rule is validated for rectangle loops demonstrating invariant circulation statistics with respect to the loop aspect ratio. Near the wall, circulation PDF in the inertial range shows a Gaussian distribution compared to the stretched tails of the PDF of outer flow, consistent with the difference between inner unifractality and outer bifractality.

Several intriguing questions arise from this work that warrant further exploration. (1) As we have seen, a transition occurs between the non-intermittent scaling near the wall and the bifractality in the outer region. Determining more precise details of this transition and examining the corresponding flow patterns and properties could enhance our understanding of intermittency in wall turbulence.
(2) While in HIT and quantum turbulence, connections between circulation and vortical structures are well studied \citep{Muller21_NC_quantum, Moriconi2022_PRE}, they remain unexplored in wall flows, and it is unclear why $\zeta_p$ deviates from K41 at $p=3$. (3) It would be very interesting to detect the circulation statistics in $x$-$y$ and $y$-$z$ planes, which contain information of spanwise and streamwise vorticity and are expected to provide quantitative insights into near-wall vortex structures. They will not have the advantage of homogeneity as for wall-parallel planes, so the boxes on which to compute the circulation have to be guided by our sense of the physics of the vertical structure of the channel flow.
(4) Finally, the relative simplicity of the scaling properties of circulation, in contrast to the multifractal nature of velocity increments, cannot be overemphasized.


\backsection[Acknowledgements]{The authors acknowledge \cite{LM2015} and the Johns Hopkins Turbulence Database \citep{JHU_database} for access to DNS data at $Re_\tau=1000$ and 5200.}

\backsection[Funding]{XC thanks for the support by the National Natural Science Foundation of China (grant numbers 92252201, 12072012) and the "Fundamental Research Funds for the Central Universities". KRS thanks New York University for support of his part of this research.
}



\backsection[Author ORCID]{
Peng-Yu Duan, https://orcid.org/0009-0002-0579-3356;
Xi Chen, http://orcid.org/0000-0002-4702-8735;
Katepalli R. Sreenivasan,
https://orcid.org/0000-0002-3943-6827.}

\bibliographystyle{jfm}
\bibliography{jfm}

\begin{thebibliography}{22}
\expandafter\ifx\csname natexlab\endcsname\relax\def\natexlab#1{#1}\fi
\def\au#1{#1} \def\ed#1{#1} \def\yr#1{#1}\def\at#1{#1}\def\jt#1{\textit{#1}}
  \def\bt#1{#1}\def\bvol#1{\textbf{#1}} \def\vol#1{#1} \def\pg#1{#1}
  \def\publ#1{#1}\def\arxiv#1{#1}\def\org#1{#1}\def\st#1{\textit{#1}}

\bibitem[Adrian(2000)]{Adrian2000}
{\sc \au{Adrian, R.}} \yr{2000} {\em Vortex Packets and the Structure of Wall
  Turbulence Extended Abstract\/},  \pg{pp. 77--77}.  \publ{Dordrecht: Springer
  Netherlands}.

\bibitem[Alhareth {\em et~al.\/}(2024)Alhareth, Mugundhan, Langley \&
  Thoroddsen]{KAUST_PoF_2024}
{\sc \au{Alhareth, Abdullah~A.}, \au{Mugundhan, Vivek}, \au{Langley,
  Kenneth~R.} \& \au{Thoroddsen, Sigurdur~T.}} \yr{2024}  \at{Turbulent
  structure and circulation inside different planar contractions}.  \jt{Physics
  of Fluids}  \bvol{36}~(7),  \pg{075129}.

\bibitem[Benzi {\em et~al.\/}(1997)Benzi, Biferale, Struglia \&
  Tripiccione]{Benzi97_PRE_circ_in_shear}
{\sc \au{Benzi, R.}, \au{Biferale, L.}, \au{Struglia, M.V.} \& \au{Tripiccione,
  R.}} \yr{1997}  \at{Self-scaling properties of velocity circulation in shear
  flows}.  \jt{Physical Review E}  \bvol{55}~(3),  \pg{3739}.

\bibitem[Cao {\em et~al.\/}(1996)Cao, Chen \& Sreenivasan]{Cao96_PRL}
{\sc \au{Cao, N.Z.}, \au{Chen, S.Y.} \& \au{Sreenivasan, K.R.}} \yr{1996}
  \at{Properties of velocity circulation in three-dimensional turbulence}.
  \jt{Physical Review Letters}  \bvol{76}~(4),  \pg{616}.

\bibitem[Frisch(1995)]{Frisch_turbulence}
{\sc \au{Frisch, Uriel}} \yr{1995} {\em Turbulence: the legacy of AN
  Kolmogorov\/}.  \publ{Cambridge university press}.

\bibitem[Iyer {\em et~al.\/}(2021)Iyer, Bharadwaj \& Sreenivasan]{Iyer21_PNAS}
{\sc \au{Iyer, K.P.}, \au{Bharadwaj, S.S.} \& \au{Sreenivasan, K.R.}} \yr{2021}
   \at{The area rule for circulation in three-dimensional turbulence}.
  \jt{Proceedings of the National Academy of Sciences}  \bvol{118}~(43),
  \pg{e2114679118}.

\bibitem[Iyer {\em et~al.\/}(2019)Iyer, Sreenivasan \& Yeung]{ISY19}
{\sc \au{Iyer, K.P.}, \au{Sreenivasan, K.R.} \& \au{Yeung, P.K.}} \yr{2019}
  \at{Circulation in high {R}eynolds number isotropic turbulence is a
  bifractal}.  \jt{Physical Review X}  \bvol{9}~(4),  \pg{041006}.

\bibitem[Jimenez(2018)]{Jimenez_2018}
{\sc \au{Jimenez, J.}} \yr{2018}  \at{Coherent structures in wall-bounded
  turbulence}.  \jt{Journal of Fluid Mechanics}  \bvol{842},  \pg{P1}.

\bibitem[Kline {\em et~al.\/}(1967)Kline, Reynolds, Schraub \&
  Runstadler]{Kline1967_coherent_struc}
{\sc \au{Kline, S.J.}, \au{Reynolds, W.C.}, \au{Schraub, F.A.} \&
  \au{Runstadler, P.W.}} \yr{1967}  \at{The structure of turbulent boundary
  layers}.  \jt{Journal of Fluid Mechanics}  \bvol{30}~(4),  \pg{741--773}.

\bibitem[Lee \& Moser(2015)]{LM2015}
{\sc \au{Lee, M.} \& \au{Moser, R.D.}} \yr{2015}  \at{Direct numerical
  simulation of turbulent channel flow up to ${R}e_\tau=5200$}.  \jt{Journal of
  Fluid Mechanics}  \bvol{774},  \pg{395--415}.

\bibitem[Li {\em et~al.\/}(2008)Li, Perlman, Wan, Yang, Meneveau, Burns, Chen,
  Szalay \& Eyink]{JHU_database}
{\sc \au{Li, Y.}, \au{Perlman, E.}, \au{Wan, M.P.}, \au{Yang, Y.K.},
  \au{Meneveau, C.}, \au{Burns, R.}, \au{Chen, S.Y.}, \au{Szalay, A.} \&
  \au{Eyink, G.}} \yr{2008}  \at{A public turbulence database cluster and
  applications to study {L}agrangian evolution of velocity increments in
  turbulence}.  \jt{Journal of Turbulence} ~(9),  \pg{N31}.

\bibitem[Migdal(1994)]{Migdal94_area_rule}
{\sc \au{Migdal, A.A.}} \yr{1994}  \at{Loop equation and area law in
  turbulence}.  \jt{International Journal of Modern Physics A}  \bvol{9}~(08),
  \pg{1197--1238}.

\bibitem[Migdal(2023)]{Migdal2023}
{\sc \au{Migdal, A.A.}} \yr{2023}  \at{Statistical equilibrium of circulating
  fluids}.  \jt{Physics Reports}  \bvol{1011},  \pg{1--117}.

\bibitem[Moriconi {\em et~al.\/}(2022)Moriconi, Pereira \&
  Valad{\~a}o]{Moriconi2022_PRE}
{\sc \au{Moriconi, L.}, \au{Pereira, R.~M.} \& \au{Valad{\~a}o, V.~J.}}
  \yr{2022}  \at{Circulation statistics and the mutually excluding behavior of
  turbulent vortex structures}.  \jt{Physical Review E}  \bvol{106}~(2),
  \pg{L023101}.

\bibitem[Mugundhan \& Thoroddsen(2023)]{KAUST_circ_JoT}
{\sc \au{Mugundhan, V.} \& \au{Thoroddsen, S.T.}} \yr{2023}  \at{Circulation in
  turbulent flow through a contraction}.  \jt{Journal of Turbulence}
  \bvol{24}~(11-12),  \pg{577--612}.

\bibitem[M{\"u}ller \& Krstulovic(2024)]{Muller24_PRL_quantum_to_2D}
{\sc \au{M{\"u}ller, N.P.} \& \au{Krstulovic, G.}} \yr{2024}  \at{Exploring the
  equivalence between two-dimensional classical and quantum turbulence through
  velocity circulation statistics}.  \jt{Physical Review Letters}
  \bvol{132}~(9),  \pg{094002}.

\bibitem[M{\"u}ller {\em et~al.\/}(2021)M{\"u}ller, Polanco \&
  Krstulovic]{Muller21_PRX_quantum}
{\sc \au{M{\"u}ller, N.P.}, \au{Polanco, J.I.} \& \au{Krstulovic, G.}}
  \yr{2021}  \at{Intermittency of velocity circulation in quantum turbulence}.
  \jt{Physical Review X}  \bvol{11}~(1),  \pg{011053}.

\bibitem[Polanco {\em et~al.\/}(2021)Polanco, M{\"u}ller \&
  Krstulovic]{Muller21_NC_quantum}
{\sc \au{Polanco, J.I.}, \au{M{\"u}ller, N.P.} \& \au{Krstulovic, G.}}
  \yr{2021}  \at{Vortex clustering, polarisation and circulation intermittency
  in classical and quantum turbulence}.  \jt{Nature Communications}
  \bvol{12}~(1),  \pg{7090}.

\bibitem[Sreenivasan \& Antonia(1997)]{Sreeni1997_ARFM_turb}
{\sc \au{Sreenivasan, K.R.} \& \au{Antonia, R.A.}} \yr{1997}  \at{The
  phenomenology of small-scale turbulence}.  \jt{Annual Review of Fluid
  Mechanics}  \bvol{29}~(1),  \pg{435--472}.

\bibitem[Sreenivasan {\em et~al.\/}(1995)Sreenivasan, Juneja \&
  Suri]{Sreeni1995_circ}
{\sc \au{Sreenivasan, K.R.}, \au{Juneja, A.} \& \au{Suri, A.K.}} \yr{1995}
  \at{Scaling properties of circulation in moderate-{R}eynolds-number turbulent
  wakes}.  \jt{Physical Review Letters}  \bvol{75}~(3),  \pg{433}.

\bibitem[Xie {\em et~al.\/}(2021)Xie, He, Bao \& Chen]{Jiabin21}
{\sc \au{Xie, J.B.}, \au{He, J.C.}, \au{Bao, Y.} \& \au{Chen, X.}} \yr{2021}
  \at{A low-communication-overhead parallel {DNS} method for the 3{D}
  incompressible wall turbulence}.  \jt{International Journal of Computational
  Fluid Dynamics}  \pg{pp. 1--20}.

\bibitem[Zhu {\em et~al.\/}(2023)Zhu, Xie \& Xia]{Zhu23_PRL_2D_circ}
{\sc \au{Zhu, H.Y.}, \au{Xie, J.H.} \& \au{Xia, K.Q.}} \yr{2023}
  \at{Circulation in quasi-2{D} turbulence: Experimental observation of the
  area rule and bifractality}.  \jt{Physical Review Letters}  \bvol{130}~(21),
  \pg{214001}.

\end{thebibliography}

\end{document}